\documentclass[twocolumn,showpacs,preprintnumbers,amsmath,amssymb,aps,dvips,prb]{revtex4}
\usepackage{amsmath}
\usepackage[dvips]{graphicx}
\usepackage{dcolumn}
\newcolumntype{d}[1]{D{.}{.}{#1}}
\usepackage{bigdelim}
\usepackage{bm}
\usepackage{array}
\usepackage{longtable}
\usepackage{lscape}
\usepackage{float}
\usepackage{rotating}
\usepackage{multirow}
\usepackage{setspace}

\makeatletter

\newcommand{\Rmnum}[1]{\expandafter\@slowromancap\romannumeral #1@}
\makeatother

\begin{document}

\preprint{APS/123-QED}

\title{Entanglement Perturbation Theory for Antiferromagnetic Heisenberg Spin Chains}
\author{Lihua Wang$^{1,2}$}
\author{Sung Gong Chung$^{1,3}$}
\affiliation{$^1$
Department of Physics and Nanotechnology Research and Computation Center, Western Michigan University, Kalamazoo, Michigan 49008, USA  
}%
\affiliation{$^2$
Computational Condensed-Matter Physics Laboratory, \\
RIKEN ASI, Wako, Saitama 351-0198, Japan}
\affiliation{$^3$
Asia Pacific Center for Theoretical Physics, Pohang, Gyeonbuk 790-784, South Korea}

\date{\today}

\begin{abstract} 
A recently developed numerical method, entanglement perturbation theory (EPT), is used to
study the antiferromagnetic Heisenberg spin chains with z-axis anisotropy $\lambda$ and magnetic field
B. To demonstrate the accuracy, we first apply EPT to the isotropic spin-$\frac{1}{2}$ antiferromagnetic
Heisenberg model, and find that EPT successfully reproduces the exact Bethe Ansatz results for
the ground state energy, the local magnetization, and the spin correlation functions (Bethe ansatz result is available for the first 7 lattice separations). In particular, EPT confirms for the first time the asymptotic behavior of the spin correlation functions predicted by the conformal field theory, which realizes only for lattice separations larger than 1000. Next, turning on the z-axis anisotropy and the magnetic field, the 2-spin and 4-spin correlation functions are calculated, and the results are compared with those obtained by Bosonization and density matrix renormalization group methods. Finally, for the spin-1 antiferromagnetic Heisenberg model, the ground state phase diagram in $\lambda$ space is determined with help of the Roomany-Wyld RG finite-size-scaling. The results are in good agreement with those obtained by the level-spectroscopy method.

\end{abstract}

\maketitle

\section{\label{sec:introductionl}Introduction}
The physics of strongly correlated systems has been one of the central interests in condensed matter physics since the discovery of high-temperature superconductivity in 1986 \cite{bed}. The successive discovery of a variety of exotic materials over the last two decades, showing superconductivity, magnetism and their co-existence, has further fuelled to our desire for the development of powerful numerical methods to study quantum many-body systems\cite{ali}. While the various kinds of mean field theory (MFT), the dynamical MFT and the density functional theory (DFT) being in the forefront \cite{kot,kim}, have been applied to many different systems to understand the strong correlation phenomena, there have also been continuing efforts concurrently for the development and refinement of non-MFT many-body methods. One of the promising movements has its origin in density matrix renormalization group (DMRG) \cite{white2,Schollwock} and continues offering its refinements and modifications.  
Namely, an observation that DMRG leads to a matrix product representation of state vectors \cite{MPS11}, the infinite time evolving block decimation (iTEBD) \cite{vidal2}, and a variant using a variational principle rather than real space renormalization group (RG) to update local wave functions \cite{ver}, 
which advanced more recently\cite{cro,mac} to an improvement over the DMRG infinite algorithm \cite{white2}. The tensor extension in 2D began in the DMRG work \cite{mae}, followed by many recent works, which primarily use RG to update local wave functions, including infinite projected entangled pair states (iPEPS) \cite{jor,xia}. 

The idea of yet another non-MFT, entanglement perturbation theory (EPT) which we employ in this paper, was essentially put forward in the analysis of the classical statistical mechanics, the 2D, 3D Ising models \cite{chung1}. There, the notion of infinite, translational invariant matrix, tensor product states, not via DMRG but from the viewpoint of singular value decomposition (see Section II), and particularly the matrix, tensor product representation of {\it operators} 
was introduced to fully implement the old, well-known tool of transfer matrix method.
In other words, the {\it translationally invariant} matrix, tensor product
representations of states and operators are the essential ingredient for EPT to be maximally efficient.
A successive introduction
of the notion of density matrix $e^{-\beta H}$ with taking a mere parameter $\beta \rightarrow 0$, 
$\beta$ is {\it not} the inverse temperature, 
reduces the quantum ground state problem to that of classical statistical mechanics \cite{Chung2,Chung3}.

What are the differences among DMRG, its variants and EPT?
They are all the same in that they are essentially variational, 
but they are all different in details, and details crucially matter here. 
All of these methods are after wave functions in the matrix, tensor product form, and the key step is how to update the local wave functions. Methods discussed in Refs. \cite{white2,vidal2,mae,jor,xia} do it by real space RG.
The variants \cite{ver,cro,mac} do it without real space RG but they keep the {\it orthogonality condition} on the local wave functions, a characteristic of DMRG, by a unitary transformation. In terms of our local wave functions, see Section II, it reads as $\sum_{s=1}^{q}(\xi^s)^{\dagger}\xi^s=1$ or $\sum_{s=1}^{q}\xi^s(\xi^s)^{\dagger}=1$.
While the variants appear to improve on DMRG in 1D, a real test would be in 2D where DMRG has not been quite successful, and where the orthogonality condition 
remains to be successfully implemented. While some progress has been made by \cite{mae,jor,xia}, 
it is quite interesting to see how far one can go along these lines in 2D.  
On the other hand, EPT treats local wave functions with neither RG nor imposing the orthogonality condition both in 1D and 2D. 
The penalty is a numerical instability that sometimes occurs in the iterative process. Now notice that EPT is based on a distinctive philosophy, 
namely, the idea of {\it divide and conquer} realized mathematically
by singular value decomposition (SVD).  SVD allows us to write everything from state vectors 
to operators in the matrix, tensor products. 
Besides, there is no reason for the orthogonality condition in EPT. Therefore, EPT is undoubtedly simplest, 
and thus has a great potential for 2D {\it only if} 
we can establish a numerically stable algorithm.
So far, we have succeeded in 1D, as will be demonstrated in this paper, quasi-1D in another paper\cite{wang2}, 
but only partially yet in 2D \cite{chung1,Chung3}.

The model system we study in this paper is the 1D antiferromagnetic (AF) Heisenberg model described by
\begin{align}
\label{hamiltonmf}
H = J\sum_{i=1}^{L}\left[\frac{1}{2}\left(S_i^+\cdot S_{i+1}^-+S_i^-\cdot S_{i+1}^+\right)+\lambda S_i^z\cdot S_{i+1}^z\right]
\end{align}
, where $S^{\pm}=S^x\pm iS^y$ are the spin flip operators. $J$ is the exchange coupling (taken to be $1$) and $\lambda$ denotes the exchange anisotropy. $\lambda=0$ corresponds to the xy model, 1 the 
xxx model, $\infty$ the Ising model and general $\lambda$ the xxz model. Despite its simplicity, the Heisenberg spin chain fascinated physicists for generations. It is now one of few many-body 
quantum systems that are understood fully, by both analytic and numerical tools. 

The first analytic breakthrough is Bethe ansatz (BA) \cite{Bethe,Hulthen,yang1,yang2,yang3}. 
It rigorously solves for the ground state (GS) and the excitation spectrum, though a 
complete equation 
is computationally challenging to solve \cite{bethe1,bethe2,bethe3}. 
Some rigorous results of BA are essential ingredients for other theories developed later such as 
bosonization \cite{Haldane1} and conformal field theory (CFT) \cite{Ginsparg}. 
Nevertheless BA has limitation to calculate the long-distance correlations \cite{Takahashi}. 
In contrast, those effective field theories emphasize on the long-distance behavior of the model.

Bosonozation transforms the xxz model into a quantum sine-Gordon model expressed in 
bosonic field operators \cite{Miranda}. In the xxx model, the so-called Umklapp scattering term
, from $S^zS^z$ interaction, vanishes marginally after iterative renormalization group transformations, with its fixed point being the Gaussian Hamiltonian \cite{Miranda,Nagaosa}. Due to a 
spin rotational symmetry, it is then shown \cite{Nagaosa,Luther1,Luther2} that the 
spin-spin correlation decays as $r^{-1}$, $r$ being the spin separation.

The CFT, on the other hand, shows that \cite{affleck1} there is a multiplicative correction, 
$\left(\mathrm{ln} r\right)^{\frac{1}{2}}$, to the spin-spin correlation functions, 
through a renormalization group analysis of the conformal dimension which receives 
correction from the (marginally irrelevant) Umklapp term. 
The exact long-distance behavior 
$\left(2\pi\right)^{-\frac{3}{2}}\left(ln r\right)^{\frac{1}{2}}r^{-1}$ is derived in \cite{affleck}.
The power of CFT lies in the highly stringent conformal invariance in the theory of
complex functions, leading to the determination of the entire spectrum solely in terms
of the central charge and confromal dimensions, the latter in turn leading to the
spin-spin correlation functions in the long distance \cite{Nagaosa}.

On the numerical side, there are a variety of works. The typical ones are 
Quantum Monte Carlo \cite{Shoudan,nightingale} and DMRG \cite{white1,white2,Hikihara}. 
Recently researchers are numerically exploring spin chains with impurities \cite{schlottmann}, the mixed spin-1-spin-$\frac{1}{2}$ chain \cite{pati} and the spin chain with anisotropic coupling strength in the x, y, z directions \cite{xxz}. In another EPT work, the EPT-e algorithm \cite{chung5}, following Feynman's line \cite{Feynman}, was reported for the excitation spectrum of the spin chains, in particular the spin triplet excitation 
spectrum of the spin-1 chain for the whole Brillouin zone including the Haldane gap at $k=\pi$. EPT-e calculations rely on EPT-g calculations to get a very accurate ground state wave function. 

After describing the detailed formalism in section II, we apply EPT to study the 
spin $\frac{1}{2}$ AF Heisenberg chains in section III, 
where we first demonstrate the accuracy of EPT in 
comparison with BA results in Sec.IIIA, and then focus on a macroscopically long-distance behavior of the spin-spin correlations, and thus examine how far one can go with the size of entanglement $p$, 
roughly a measure of correlations. In this regard, we could go as far as $p=209$ 
(main time consumption in this calculation is taken by an eigenvalue decomposition of matrices 
with rank of $4p^2$ and a generalized eigenvalue equation of matrices with rank of $2p^2$. 
See the definition of the rank of matrices in section II) for the spin-$\frac{1}{2}$ xxx model. 
In Sec.IIIB, Bosonization's prediction on the anisotropic Heisenberg chain 
in a finite external magnetic field is confirmed. In particular, EPT has turned out to 
be more accurate than DMRG in the 4-spin correlation functions. 
However the critical exponent given by Bosonization for the xxx model is found to be 
slightly different from the exact result.  Thus another field thoery, CFT, is examined in Sec.IIIC.
Its prediction on the asymptotic behavior, which is found to start around 1000 in lattice separations, is confirmed for the first time to our best knowledge. Next, in section IV, we study the ground state phase diagram of the spin-1 xxz AF Heisenberg chain, combining EPT with 
the Roomany-Wyld RG-finite-size-scaling. We can still use EPT-g algorithm,  
that is originally designed for the ground state, to look for the first excited state 
of this spin chain.  This is based on the fact that the first excited state is at 
the wave-number $k=\pi$, which can be handled by the bipartite local wave functions 
$\xi^{1,2}$ employed in EPT-g. Our result agrees with the phase diagram obtained 
by the level-spectroscopy method \cite{wei}. Finally, we conclude the paper with discussions.
  
\section{\label{sec:ept}EPT}

EPT first introduces the local wave functions and then couples them by the entanglement to represent the translationally symmetric wave function. Second it reaches a general yet simple algebraic procedure to solve for the local wave functions, during which the uniform EPT will utilize the translational symmetry of the Hamiltonian to greatly simplify the formulation. The first step is common to all the EPT algorithms, therefore we will discuss it first. In fact, the resultant system wave function in EPT expressed by a successive product of the local wave functions precisely falls into the matrix product state concept \cite{MPS11}.  As for the second step, we have different ways to handle the operator, i.e., we either deal with the Hamiltonian itself or the density matrix, $e^{-\beta H}$. It is how to efficiently find the local wave functions that distinguishes EPT from other MPS methods \cite{ver,mac,MPS2, MPS3,MPS4,cro,MPS6}. Because at the beginning of EPT development it was the second method, i.e. dealing with the density matrix, that was applied to the 2D\&3D Ising models, it is called EPT-g1. The EPT-g2 algorithm was developed later to handle the Hamiltonian directly. In fact, EPT-g2 formulation is more lengthy. So we will discuss EPT-g1 first in the following Sec.\ref{EPT-g1} and then EPT-g2 in Sec.\ref{EPT-g2}. The comparison between them is given in Sec.\ref{EPTGSUM}.  

\begin{figure}
\begin{center}
\includegraphics[height=2.8in, width=3.2in]{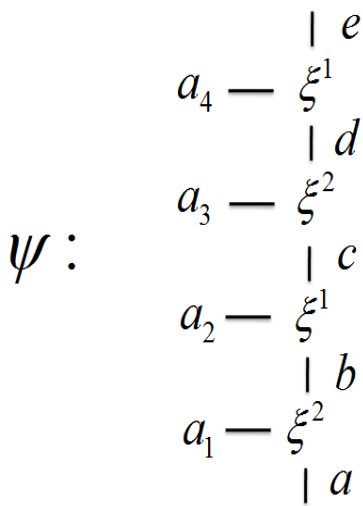}
\caption{\label{wavefig1}Schematic figure of the wave function}
\end{center}
\end{figure}
We now take a 4-sites spin chain as an example to illustrate how to write the system wave function 
in terms of local wave functions. The number of local spin states is denoted as $q$. So $q$ is 2 for 
spin-$\frac{1}{2}$ and 3 for spin-1. The wave function of this spin chain is written as $\psi\left(a_1,a_2,a_3,a_4\right)$, which is a $q^4$ dimensional vector. However it can also be regarded as a $q$ x $q^3$ matrix $\psi\left(a_1,a_2a_3a_4\right)$. 
We then use singular value decomposition (SVD) to {\it separate} $a_1$ from the rest as  

\begin{align}
\psi\left(a_1,a_2a_3a_4\right) & = A_{a_1,a}\lambda_a B_{a_2a_3a_4,a}\notag\\
& = A'_{a_1,a}B_{a_2a_3a_4,a}\notag
\end{align}
where and below the repeating index means summation. The eigenvalue $\lambda_a$ is absorbed into $A_{a_1,a}$ 
to give $A'_{a_1,a}$. We then SVD $B_{a_2a_3a_4,a}$ to {\it separate} $a_2$ from $a_3a_4$, we have
\begin{align}
B_{a_2a_3a_4,a} & = C_{a_2,ab}\eta_bD_{a_3a_4,b}\notag\\
& = C'_{a_2,ab}D_{a_3a_4,b}\notag
\end{align}
We keep using SVD until all the lattice sites are separated from each other and the eigenvalues are properly absorbed into the corresponding left matrices, which can now be regarded as the wave functions of local states $a_1,a_2,a_3,a_4$. 
Finally the system wave function ends up as
\begin{equation}
\label{wavesvd}
\psi\left(a_1,a_2,a_3,a_4\right) = A'_{a_1,a}C'_{a_2,ab}E'_{a_3,bc}F_{a_4,c}
\end{equation}
 
The above shows how to use SVD to rewrite the wave function as a successive product of local ones. 
We borrow the terminology 'entanglement' from quantum information theory to denote the coupling 
indices $a,b,c,d$ in equation (\ref{wavesvd}). Now recall that the spin chain is translational 
symmetric and has a periodic boundary condition. We should introduce some uniform local 
wave function $\xi$ to represent the  vectors $A',C',E',F$. Considering the possible $\uparrow\downarrow$ 
or $\downarrow\uparrow$ spin arrangement of AF spin chains, we need two 
local wave functions $\xi^1$ and $\xi^2$ to include such a bipartite structure. Therefore we have the SVD-ed wave function for a general spin chain as follows
\begin{equation}
\label{wavefunction1}
\psi=\cdots\xi^2_{aba_1}\otimes\xi^1_{bca_2}\otimes\xi^2_{cda_3}\otimes\xi^1_{dea_4}\cdots
\end{equation}
where, the subscript 1 on the shoulder refers to the odd sites of the chain while 2 the even sites. The indices 
$\{a_1,a_2,a_3,a_4\}$ account for the local spin states, running from 1 to 2 for spin-$\frac{1}{2}$, 1 to 3 for spin-1. 
The indices $\{a,b,c,d,e\}$ are the entanglement indices, which are integers, uniformly coupling the local 
wave functions over all spin sites. Fig.\ref{wavefig1} illustrates how the system wave function is 
expressed as a successive product of the local wave functions.

Given the system wave function in the form of (\ref{wavefunction1}), we show how to derive the main equation to solve for $\xi^1$ and $\xi^2$ in general. First recall the way to derive the Schr\"{o}dinger equation 
by the variational principle, where one varies the expectation value of the Hamiltonian (energy) 
with respect to the system wave function. In the algebraic form, the Schr\"{o}dinger equation is an eigenvalue problem. One now has to similarly vary the expectation value of a functional of the Hamiltonian with respect to, for example, $\xi^1$ as follows.   
\begin{equation}
\label{maineqn}
\frac{\delta}{\delta \xi^1} \left(\frac{\langle \psi\left(\xi^1,\xi^2\right)\mid \textit{f}\left( H\right)\mid \psi\left(\xi^1,\xi^2\right) \rangle}{\langle \psi\left(\xi^1,\xi^2\right) \mid \psi\left(\xi^1,\xi^2\right) \rangle}\right)=0
\end{equation}

We thus arrive at a generalized eigenvalue equation,

\begin{align}
\label{eptgeigen}
X_i\left(\xi^1,\xi^2\right)\xi^i & = \mu Y_i\left(\xi^1,\xi^2\right)\xi^i \hspace{10 mm} i=1,2
\end{align}

The GS of the system is corresponding to either the smallest or the largest eigenvalue of (\ref{eptgeigen}), depending on the functional $\textit{f}\left(H\right)$. We see that it is possible to make (\ref{eptgeigen}) self-consistent by starting with two vectors $\xi_0^1$ and $\xi_0^2$ as a seed, and to determine $X_1$ and $X_2$, hence to determine a set of $\xi_1^1$ and $\xi_1^2$ as the new seed. 
The iteration is continued this way until the convergence is reached. Note that the 
size of the generalized eigenvalue problem is controlled by the entanglement, $p$. 
More precisely, it is $2p^2$ for spin-$\frac{1}{2}$ and $3p^2$ for spin-1. The larger entanglement will give 
the more precise representation of the wave function. If it is large enough, the wave function 
could be exact. So we start with a small entanglement, say 1, to get the converged $\xi^1$ and $\xi^2$ 
for it. Then we increase the entanglement and solve for the corresponding $\xi$'s. We compare the physical 
quantities calculated from the local wave functions with increasing entanglements to seek for the 
convergence. EPT calculations show that the entanglement necessary for the convergence increases 
slower than a linear function of the system size and that the convergence for each entanglement 
is rather quick. Therefore it has high efficiency, which enables one to calculate very 
large system precisely because now the size to be handled is roughly of order $L^2$ vs. $q^L$ in 
the exact diagonalization. In short, EPT is a method of {\it divide} and {\it conquer} realized by SVD, Eq.(3) $=$ 
{\it divide} and the generalized eigenvalue equation, Eq.(5) $=$ {\it conquer}.

\subsection{\label{EPT-g1}EPT-g1}

Since any eigenstate $\mid \psi_i\rangle$ of $H$ is also the eigenstate of density matrix 
$e^{-\beta H}$, the GS now bears the largest eigenvalue $e^{-\beta E_0}$. 
The eigenvalue problem we try to solve in EPT-g1 is,
\begin{equation}
\label{eptg1eigenproblem1}
e^{-\beta H}\mid \psi_i\rangle=e^{-\beta E_i}\mid \psi_i\rangle
\end{equation} 
Let us consider for an illustration purpose the xxx Heisenberg Hamiltonian, and express it 
 as the sum of the local bond Hamiltonian between two nearest neighbors
\begin{equation}
\label{eptg1localhamiltonian1}
H=\sum_{bond}{H_{bond}}
\end{equation}
with
\begin{equation}
\label{eptg1localhamiltonian2}
H_{bond}\equiv \frac{J}{2}\left(S_i^{+}S_{i+1}^{-}+S_i^{-}S_{i+1}^{+}\right)+JS_i^zS_{i+1}^z
\end{equation}
We then choose an appropriate small positive $\beta$ ($0<\beta\ll1$) to safely separate the even and odd sub lattices in the density matrix
\begin{align}
\label{eptg1speratesublattice}
e^{-\beta H} & =\prod_{bond}{e^{-\beta H_{bond}}}+{\mathcal{O}}\left(\beta^2\right)\notag\\
& \approx e^{-\beta\sum_{even}{H_{bond}}}e^{-\beta\sum_{odd}{H_{bond}}}
\end{align}
Now let us linearize $e^{-\beta H_{bond}}$, again, due to the fact that $0<\beta\ll1$
\begin{align}
\label{eptg1linearize}
e^{-\beta H_{bond}} & \approx 1-\frac{J\beta}{2}\left(S_i^{+}S_{i+1}^{-}+S_i^{-}S_{i+1}^{+}\right)-J\beta S_i^zS_{i+1}^z\notag\\
 & \equiv \Omega_{\alpha}\otimes\Theta_{\alpha}
\end{align}
where $\Omega_{\alpha}$ takes four operators $1,S_i^+,S_i^-,S_i^z$ on site $i$ and $\Theta_{\alpha}$ 
likewise operators on site $i+1$. The index $\alpha$ runs from 1 to 4. The local density matrix can be written as
\begin{equation}
\label{eptg1matrix}
\langle lk\mid e^{-\beta H_{bond}}\mid ij\rangle\approx f_{\alpha,ik}\otimes g_{\alpha,jl}
\end{equation}
where, for a spin-$\frac{1}{2}$ chain,
\begin{eqnarray}
\label{operators1}
f_1 &=&  g_1 = \left( \begin{array}{ccc}
1 & 0 \\
0 & 1 \\
\end{array} \right)\\
f_2 &=& i\sqrt{\frac{\beta J}{2}}\left( \begin{array}{ccc}
0 & 1 \\
0 & 0 \\
\end{array} \right)\\
g_2 &=& i\sqrt{\frac{\beta J}{2}}\left( \begin{array}{ccc}
0 & 0 \\
1 & 0 \\
\end{array} \right)
\end{eqnarray}
and so on. For a spin-1 chain, $f$'s and $g$'s should be 3 x 3 matrices.

Now let us write down the matrix representation of the even and odd bonds in the density matrix as follows
\begin{align}
\cdots f_{\alpha}\otimes & g_{\alpha} \otimes f_{\gamma} \otimes g_{\gamma}\cdots\notag\\
\cdots & f_{\beta}\otimes g_{\beta} \otimes f_{\delta} \otimes g_{\delta}\cdots\notag
\end{align}
The vertical alignment above means that the matrix representations are at the same site. The whole density matrix will be written as
\begin{align}
\label{eptg1wholebonds1}
K & \equiv \cdots g_{\alpha}\cdot f_{\beta}\otimes f_{\gamma}\cdot g_{\beta} \cdots\notag\\
& \equiv \cdots \Gamma_{\alpha\beta}^2\otimes\Gamma_{\beta\gamma}^1\cdots
\end{align}
\begin{figure}
\begin{center}
\includegraphics[height=2.8in, width=3.2in]{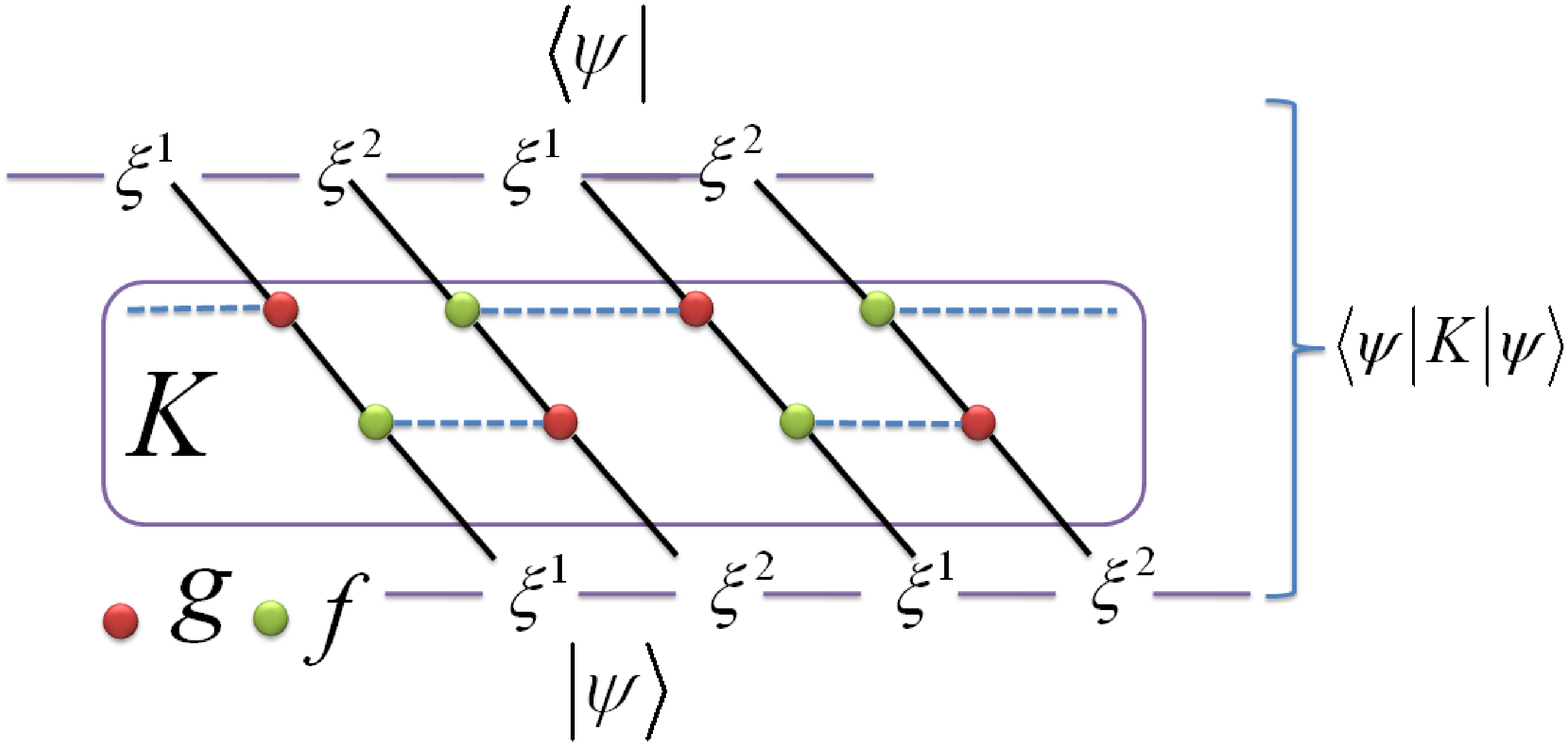} 
\caption{\label{eptg1transferfig}Schematic figure of expectation value of the density matrix in EPT-g1}
\end{center}
\end{figure}
Fig.\ref{eptg1transferfig} illustrates the expectation value of the density matrix. 

\begin{figure}
\begin{center}
\includegraphics[height=2.8in, width=3.2in]{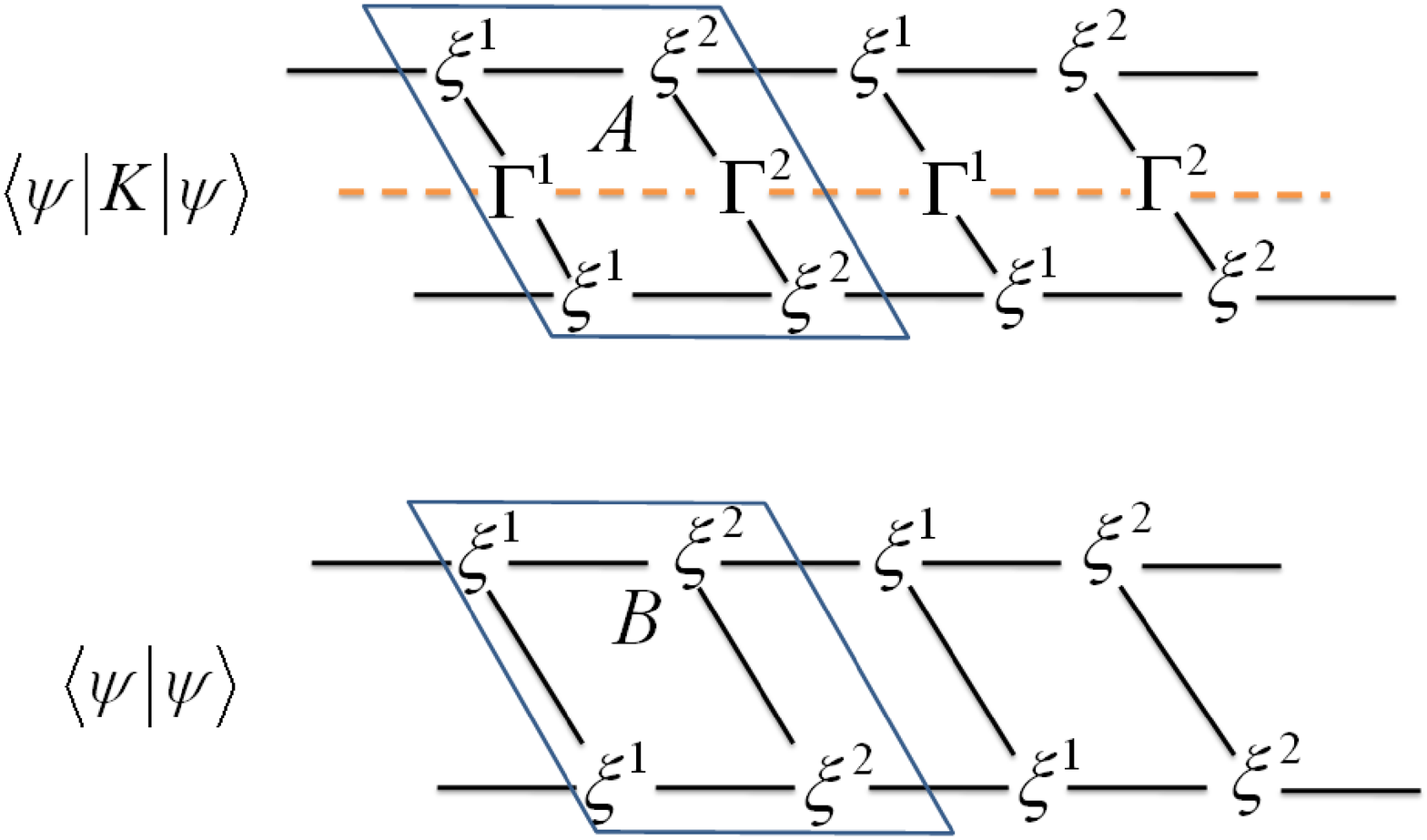} 
\caption{\label{eptg1formab}Schematic figure of forming the translationally symmetric units in EPT-g1}
\end{center}
\end{figure}

$K$ is now substituted into (\ref{maineqn}) as the functional $\textit{f}$ to have
\begin{align}
\langle\psi \left(\xi^1,\xi^2 \right)\mid K\mid\psi\left(\xi^1,\xi^2\right)\rangle & = Tr\left(A^{L/2}\right) \\
\langle\psi \left(\xi^1,\xi^2 \right)\mid \psi\left(\xi^1,\xi^2\right)\rangle & = Tr\left(B^{L/2}\right)
\end{align}
where the matrices $A$ and $B$ are the translationally symmetryic units illustrated in Fig.\ref{eptg1formab}. Explicitly
\begin{align}
\label{eptg1matrixAB}
A_{m_1,n_1} & = \xi_{aea_1}^1\xi_{eca_2}^2\xi_{bfb_1}^1\xi_{fdb_2}^2\Gamma^1_{\alpha\beta,a_1a_2}\Gamma^2_{\beta\gamma,b_1b_2}\\
B_{m,n} & = \xi_{aea_1}^1\xi_{eca_2}^2\xi_{bfa_1}^1\xi_{fda_2}^2\\
m_1 & = (\gamma-1)\times p^2+(b-1)\times p +a\notag\\
n_1 & = (\alpha-1)\times p^2+(d-1)\times p +c\notag\\
m & = (b-1)\times p +a\notag\\
n & = (d-1)\times p +c\notag
\end{align}

Above {$a,b,c,d,e,f$} run from 1 to $p$, {$a_1,a_2,b_1,b_2$} run from 1 to $q$ and {$\alpha,\beta,\gamma$} run from 1 to 4. 
We repeat that $p$ is the entanglement and $q$ is 2 for spin-$\frac{1}{2}$, 3 for spin-1. We see that $A$ is a $4p^2$ x $4p^2$ 
matrix for spin-$\frac{1}{2}$ and $B$ is a $p^2$ x $p^2$ matrix for both spin-$\frac{1}{2}$ and spin-1. 
The variation (\ref{maineqn}) leads to
\begin{equation}
\label{eptg1aviation2}
A^{L/2-1}\ast\left(\delta A\right) = \mu B^{L/2-1}\ast\left(\delta B\right)
\end{equation}
The symbol $\ast$ denotes matrix product and then take a trace.
We then diagonalize the matrices $A$ and $B$:
\begin{align}
A=&R\Lambda L^{\dagger}\\
B=&R'\Lambda' L'^{\dagger}
\end{align}
where $R$, $\Lambda$, and $L$ are respectively the right eigenvector matrix, the diagonal eigenvalue
matrix and the left eigenvector matrix. Equation (\ref{eptg1aviation2}) is then further rewritten as follows
\begin{equation}
\label{eptg1variation3}
\left(\Lambda_i\right)^{L/2-1}L_i\left(\delta A\right)R_i\ =\mu\left(\Lambda'_j\right)^{L/2-1}L'_j\left(\delta B\right)R'_j\
\end{equation}
where $i$ runs from 1 to $4p^2$ and $j$ runs from 1 to $p^2$. Now we can explicitly work out the variation 
with respect to the local wave functions, for example, $\xi^1$. Using equation (\ref{eptg1matrixAB}), (\ref{eptg1variation3}) 
leads to the detailed form of one of the generalized eigenvalue equation (\ref{eptgeigen}),
\begin{equation}
\label{eptg1geg1}
X_1(aea_1,bfa_2)\xi^1_{bfa_2}=\mu Y_1(aea_1,bfa_2)\xi^1_{bfa_2}
\end{equation} 
where
\begin{align}
& X_1(aea_1,bfa_2)\notag\\
&=\xi^2_{eca_3}\xi^2_{fda_4}\Gamma^1_{\alpha\beta,a_1a_2}\Gamma^2_{\beta\gamma,a_3a_4}R_{cd\gamma,i}L_{ab\alpha,i}\left(\Lambda_i\right)^{L/2-1}\\
&Y_1(aea_1,bfa_2)\notag\\
&=\xi^2_{eca_3}\xi^2_{fda_3}R'_{cd,j}L'_{ab,j}\left(\Lambda'_j\right)^{L/2-1}\delta_{a_1a_2}
\end{align}
Likewise, we get the generalized eigenvalue equation for $\xi^2$. We solve for $\xi^1$ and $\xi^2$ as stated in Sec.\ref{sec:ept}.
We have next step $\xi^{1,2}$ and repeat calculation until convergence, $\xi_{old}^{1,2}=\xi_{new}^{1,2}$.

\subsection{\label{EPT-g2}EPT-g2}

EPT-g2 handles the Hamiltonian directly. The GS energy and first excited state energy can be solved by looking for the first and second minimum of the quantity
\begin{equation}
\label{energ}
\epsilon\equiv\frac{\langle\psi\mid H\mid\psi\rangle}{\langle\psi\mid\psi\rangle}
\end{equation}

Now since both the wave function and the Hamiltonian are translationally symmetric, we rewrite (\ref{energ}) as follows 
\begin{equation}
\label{energ1}
\epsilon=\frac{L}{2}\frac{\langle\psi\mid H_{bond}^{odd}\mid\psi\rangle}{\langle\psi\mid\psi\rangle}+\frac{L}{2}\frac{\langle\psi\mid H_{bond}^{even}\mid\psi\rangle}{\langle\psi\mid\psi\rangle}
\end{equation}
We have the matrix representation
\begin{equation}
\Gamma_{b_1b_2,a_1a_2}=\langle b_1\mid \langle b_2\mid H_{bond}\mid a_1\rangle\mid a_2\rangle
\end{equation}
where $\mid a_1\rangle$ and $\mid a_2\rangle$ refer to the local basis vectors on two neighboring sites while $\langle b_1\mid$ and $\langle b_2\mid$ are the conjugate counter parts. Now we are allowed to write (\ref{energ1}) further as follows
\begin{equation}
\label{energ2}
\epsilon=\frac{L}{2}\frac{Tr(A^{L/2-1}B)}{Tr(A^{L/2})}+\frac{L}{2}\frac{Tr(C^{L/2-1}D)}{Tr(A^{L/2})}
\end{equation}
where
\begin{align}
\label{matrixAB1}
A_{m,n} & =\xi_{aea_1}^1\xi_{eca_2}^2\xi_{bfa_1}^1\xi_{fda_2}^2\\
B_{m,n} & = \xi_{aea_1}^1\xi_{eca_2}^2\xi_{bfb_1}^1\xi_{fdb_2}^2\Gamma_{b_1b_2,a_1a_2}\\
m & = (b-1)\times p +a\notag\\
n & = (d-1)\times p +c\notag
\end{align}

Above {$a,b,c,d,e,f$} run from 1 to $p$ and {$a_1,a_2,b_1,b_2$} run from 1 to $q$. We can write down the matrix 
C and D likewise. 
\begin{figure}
\begin{center}
\includegraphics[height=2.8in, width=3.4in]{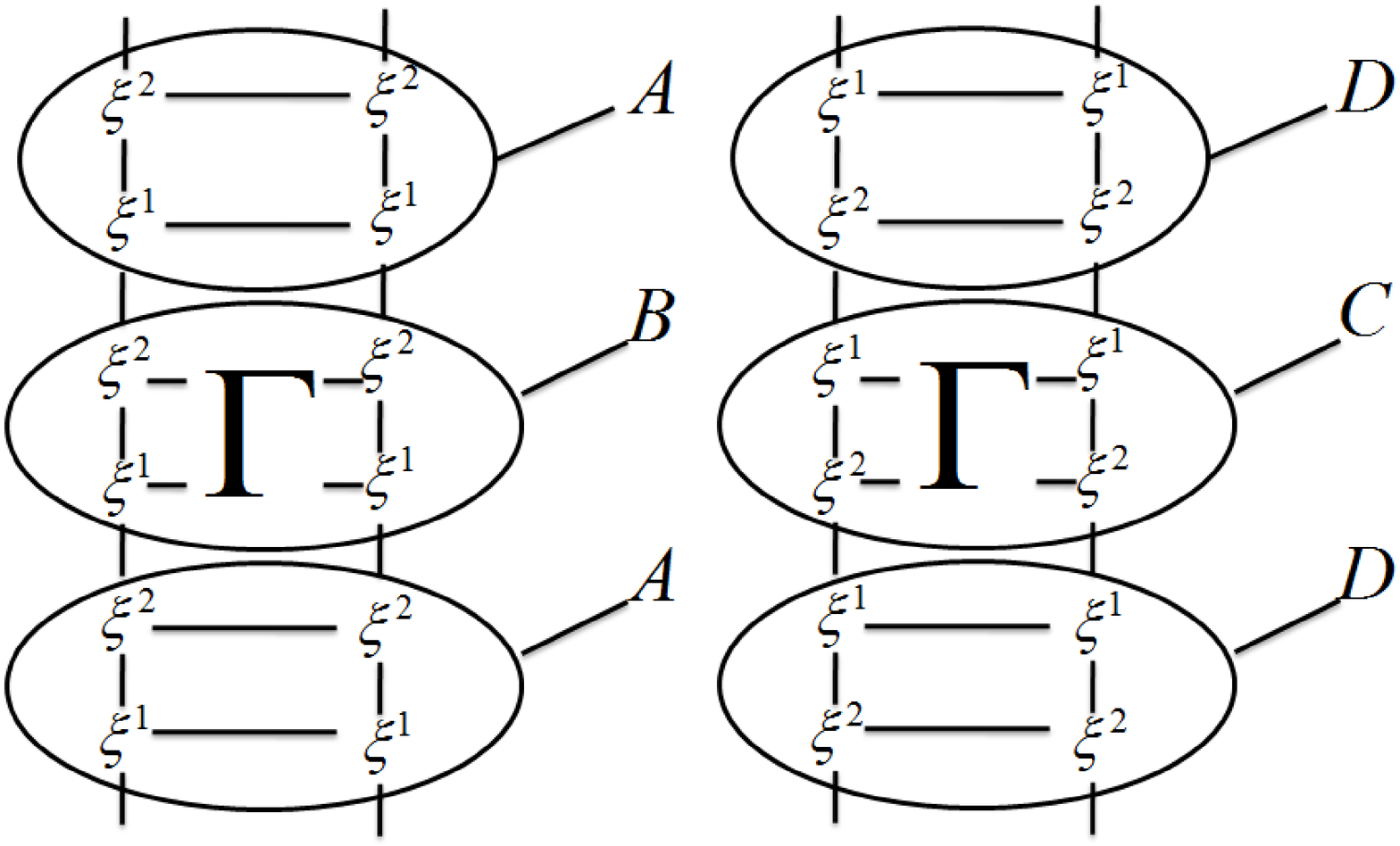}
\caption{\label{fig:sandwitch0}Schematic figure of the formation of matrixes A, B, C and D in EPT-g2}
\end{center}
\end{figure}
Fig.\ref{fig:sandwitch0} schematically shows the formation of matrixes A, B, C and D.  
If we vary (\ref{energ2}) with respect to $\xi^1$ and $\xi^2$, we will arrive at
\begin{align}
\label{aviation1}
& \epsilon A^{L/2-1}\ast\left(\delta A\right)\notag\\
 = & \left(\delta B\right)\ast A^{L/2-1}+\sum_{j=1}^{L/2-1}B\ast A^{L/2-j-1}\ast\left(\delta A\right)\ast A^{j-1}\notag\\
& +\left(\delta D\right)\ast C^{L/2-1}+\sum_{j=1}^{L/2-1}D\ast C^{L/2-j-1}\ast\left(\delta C\right)\ast C^{j-1}
\end{align}
We then diagonalize the matrices $A$ and $C$:
\begin{align}
A=&R\Lambda L^{\dagger}\\
C=&R'\Lambda' L'^{\dagger}
\end{align}
Equation (\ref{aviation1}) is rewritten as follows
\begin{align}
\label{aviation2}
& \epsilon\sum_{i=1}^{P^2}\left(\Lambda_i\right)^{L/2-1}L_i\left(\delta A\right)R_i\notag\\
= & \sum_{i=1}^{P^2}{L_i\left(\Lambda_i\right)^{L/2-1}\left(\delta B\right)R_i}+\sum_{i=1}^{P^2}{L'_i\left(\Lambda'_i\right)^{L/2-1}\left(\delta D\right)R'_i}\notag\\
& +\textit{f}_1\left(L,R,\Lambda,\delta A\right)+\textit{f}_2\left(L',R',\Lambda',\delta C\right)
\end{align}
where
\begin{widetext}
\begin{align}
\textit{f}_1\left(L,R,\Lambda,\delta A\right)& = \sum_{m,n=1;m\ne n}^{P^2}{\frac{\Lambda_m^{L/2-1}-\Lambda_n^{L/2-1}}{\Lambda_m-\Lambda_n}}E_{mn}L_n\left(\delta A\right)R_m +
(L/2-1)\sum_{m=1}^{P^2}{\Lambda_m^{L/2-2}}E_{mm}L_m\left(\delta A\right)R_m\notag\\
\textit{f}_2\left(L',R',\Lambda',\delta C\right) & = \sum_{m,n=1;m\ne n}^{P^2}{\frac{\Lambda_m^{'L/2-1}-\Lambda_n^{'L/2-1}}{\Lambda'_m-\Lambda'_n}F_{mn}L'_n\left(\delta C\right)R'_m}+
(L/2-1)\sum_{m=1}^{P^2}{\Lambda_m^{'L/2-2}F_{mm}L'_m\left(\delta C\right)R'_m}\notag\\
E_{mn} & = L_mBR_n\notag\\
F_{mn} & = L'_mDR'_n\notag
\end{align}
\end{widetext}
Like in EPT-g1, we solve (\ref{eptgeigen}) and hence the problem. Note that the smallest eigenvalue 
and corresponding eigenvector give the GS energy and the local wave functions. 
The second smallest will bring about the first excited state energy and the 
corresponding wave functions as long as the excited state has an appropriate wave function structure.

\subsection{\label{EPTGSUM}Summary}

In the last two subsections, we saw that both EPT-g1 and EPT-g2 can solve for the GS. 
The EPT-g1 is simpler, but the calculation is heavier because the matrix $A$ for 
EPT-g1 is larger than $B$ and $D$ for EPT-g2. However, EPT-g1 is especially suitable 
for infinite systems with translation symmetry, because only the largest eigenvalue 
of $A$ needs be retained. On the other hand, EPT-g2 has matrices $E$ and $F$ containing 
all the eigenvalues and eigenvectors of $A,B,C,D$, hence it becomes slower than EPT-g1 
for infinite systems. Actually we used both EPT-g1 and EPT-g2 ($\beta$ is set to be 
$10^{-6}$ for up to thousand spin sites) for finite spin chains. Since EPT-g2 does not 
make any approximation and the two methods give exactly the same results and converge 
with the entanglement at the same speed, the small parameter $\beta$ in EPT-g1 algorithm 
is not an issue. Later we will see EPT-g1 calculation for infinite spin-$\frac{1}{2}$ chains gives 
very accurate results. Moreover EPT-g's can be applied to the first excited state for 
spin-1 chains as well, because the first excitation occurs when the wave 
number $k=\pi$ which means the bipartite trial wave function is suitable. On the other hand, 
the first excited state in spin-$\frac{1}{2}$ chains cannot be calculated by EPT-g's since it does not 
have this character. We then need to use EPT-e as discussed in \cite{chung5}.

\section{Spin-$\frac{1}{2}$ chains}

First we use EPT-g to solve for the GS  
without the magnetic field, and calculate the energy, the local 
magnetization and spin-spin correlations. We especially investigate the long-range 
behavior of 2-spin correlations, considering a very 
large entanglement and the infinite size, and examine the prediction of CFT \cite{affleck}. 
As mentioned in Sec.\ref{EPTGSUM}, EPT-g2 works as well as EPT-g1 for finite systems. 
But for infinite systems, EPT-g1 is more suitable.  
Second we calculate the model with z-axis anisotropy and in the external magnetic field. In particular, 
we studied both the long-range 2-spin and 4-spin correlations, to check the predictions 
of the bosonization method \cite{Hikihara}.   

\begin{table}
\begin{center}
\begin{tabular}{|c|c|c|c|}
  \hline
  chain length  & EPT result & Bethe Ansatz result \\
  \hline
  16 & -0.4463935 & -0.4463935 \\
  \hline
  64 & -0.4433459 & -0.4433485 \\
  \hline 
  256 & -0.4431551 & -0.4431597 \\
  \hline    
\end{tabular}
\caption{\label{table:ground}Comparison of the GS energies between Bethe Ansatz and EPT}
\end{center}
\end{table}

\begin{figure}
\begin{center}
\includegraphics[height=2.8in, width=3.2in]{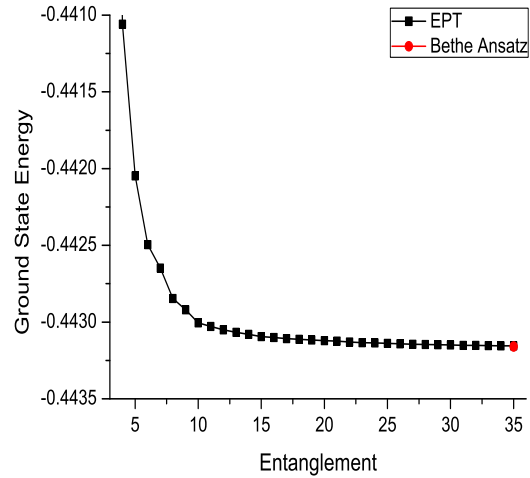}
\caption{\label{fig:converg}Convergence with entanglement of the GS energy for a 256-sites spin-$\frac{1}{2}$ chain}
\end{center}
\end{figure}

\begin{figure}
\begin{center}
\includegraphics[height=2.8in, width=3.2in]{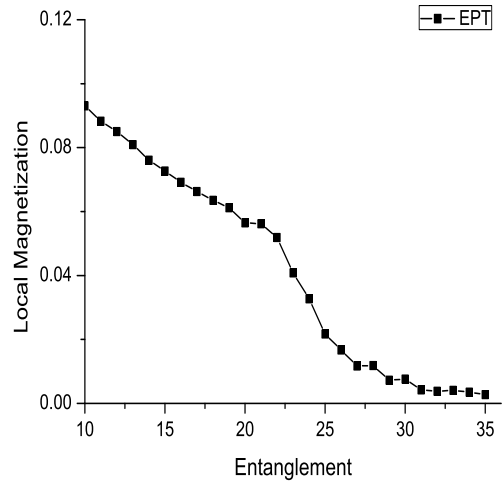}
\caption{\label{fig:mag}Convergence with entanglement of the local magnetization 
for a 256-sites spin-$\frac{1}{2}$ chain}
\end{center}
\end{figure}

\begin{figure}
\begin{center}
\includegraphics[height=2.8in, width=3.2in]{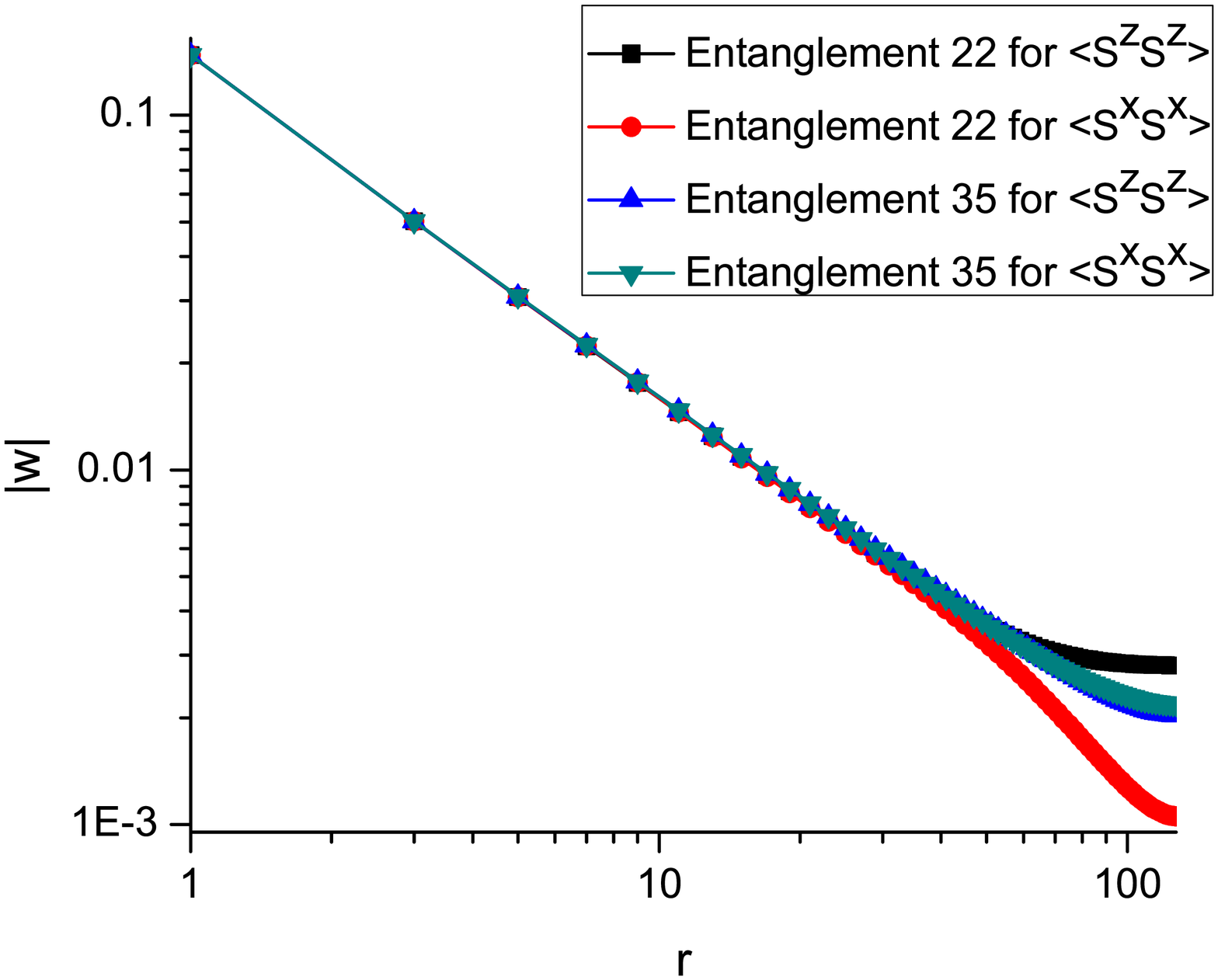}
\caption{\label{fig:corr256}Convergence with entanglement of the spin-spin correlations for a 256-sites spin-$\frac{1}{2}$ chain}
\end{center}
\end{figure}

One of the merits of EPT is that it can give precise local wave functions from which many quantities can be calculated. However, one must ask how to judge the local wave function's convergence. We can choose one of those quantities as a convergence indicator. The convergence speeds of the GS energy and the local magnetization for a 256-sites chain are shown respectively in Fig.\ref{fig:converg} 
and Fig.\ref{fig:mag}. We see that the convergence of the GS energy is reached before entanglement $p=25$ and the 
local magnetization is converged after $p=30$. Fig.\ref{fig:corr256} shows the convergence of another 
important quantity for the same chain: the spin-spin correlations, $W_{\sigma}(r)=\langle S_0^{\sigma}S_r^{\sigma}\rangle$. 
It is shown that $W_z$ and $W_x$ curves start 
to coincide after $p=35$, which is a sign of the convergence in 2-spin correlations because the symmetry among x, y and z axis are recovered. In short, the convergence of the correlation functions appears to be the ultimate indicator for 
the convergence of all quantities.   

\subsection{\label{onehalf}Comparison with Bethe ansatz}

We have calculated the spin-$\frac{1}{2}$ chains of length 16, 32, 64, 128, 256 and 512. 
EPT agrees well with Bethe ansatz \cite{bethe2} for the GS energies, as shown in Tab.\ref{table:ground}. 


To check the accuracy of the calculated correlation functions, we compare in Tab.\ref{table:corr} EPT with the Bethe ansatz (generating function method) \cite{Takahashi} for the first 7 lattice separations, which are the only spin-spin correlations handled by Bethe ansatz up to date. 
\begin{table}
\begin{center}
\begin{tabular}{|c|c|c|c|}
  \hline
  distance  & EPT 256-site & EPT infinite &Bethe Ansatz  \\
  \hline
  1 & -0.1477158 & -0.1476404 &-0.1477157 \\
  \hline
  2 & 0.0606715 & 0.0606355 &0.0606798 \\
  \hline 
  3 & -0.0502196 & -0.0501772 &-0.0502486 \\
  \hline 
  4 & 0.0346027 & 0.0346248 &0.0346528 \\
  \hline  
  5 & -0.0308457 & -0.0308430 &-0.0308904 \\
  \hline       
  6 & 0.0243621 & 0.0244263 &0.0244467 \\
  \hline  
  7 & -0.0223980 & -0.0224636 &-0.0224982 \\
  \hline  
 
\end{tabular}
\caption{\label{table:corr}Comparison of the spin-spin correlations between EPT and Bethe Ansatz over the first 7 sites}
\end{center}
\end{table}
We see that EPT agrees with BA. Note that at entanglement 180, in the third column of Tab.\ref{table:corr}, 
EPT for an infinite chain achieves the relative error of $0.1\%$ throughout the first 7 lattice separations.

\subsection{\label{onehalf1}Comparison with Bosonization}

\begin{figure}
\begin{center}
\includegraphics[height=2.8in, width=3.2in]{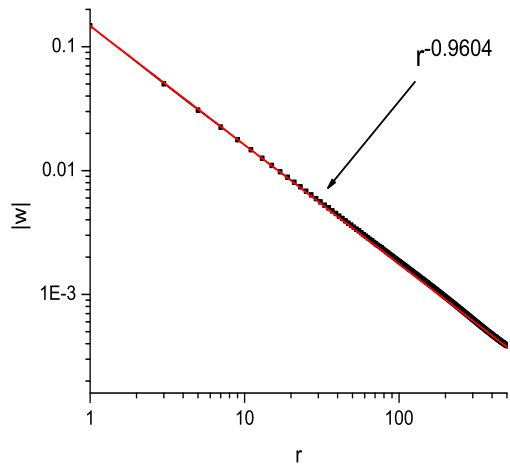}
\caption{\label{fig:fitcorrelation}Log-log plot of the spin-spin correlation functions over the 
first 501 sites for an infinite spin-$\frac{1}{2}$ chain at entanglement 180}
\end{center}
\end{figure}

Let us compare Bosonization's prediction on the spin-spin correlations with our result. 
In fig.\ref{fig:fitcorrelation} the linear fit of the log-log plot of the spin-spin 
correlations for odd separations shows a power-law decay, $W_z(r)=-0.1473 r^{-0.9604}$ 
with a fitting error of $0.1\%$. This fitting also applies to the even 
$r$ but with a positive sign reflecting the AF nature. We see that the critical exponent 
given by EPT is $-0.9604$, indeed very close to Bosonization's prediction of $-1$
but slightly different. A better prediction with a multiplicative logarithmic correction 
was given by CFT, which will be discussed in Sec.IIIC.

\begin{figure}
\begin{center}
\includegraphics[height=2.8in, width=3.2in]{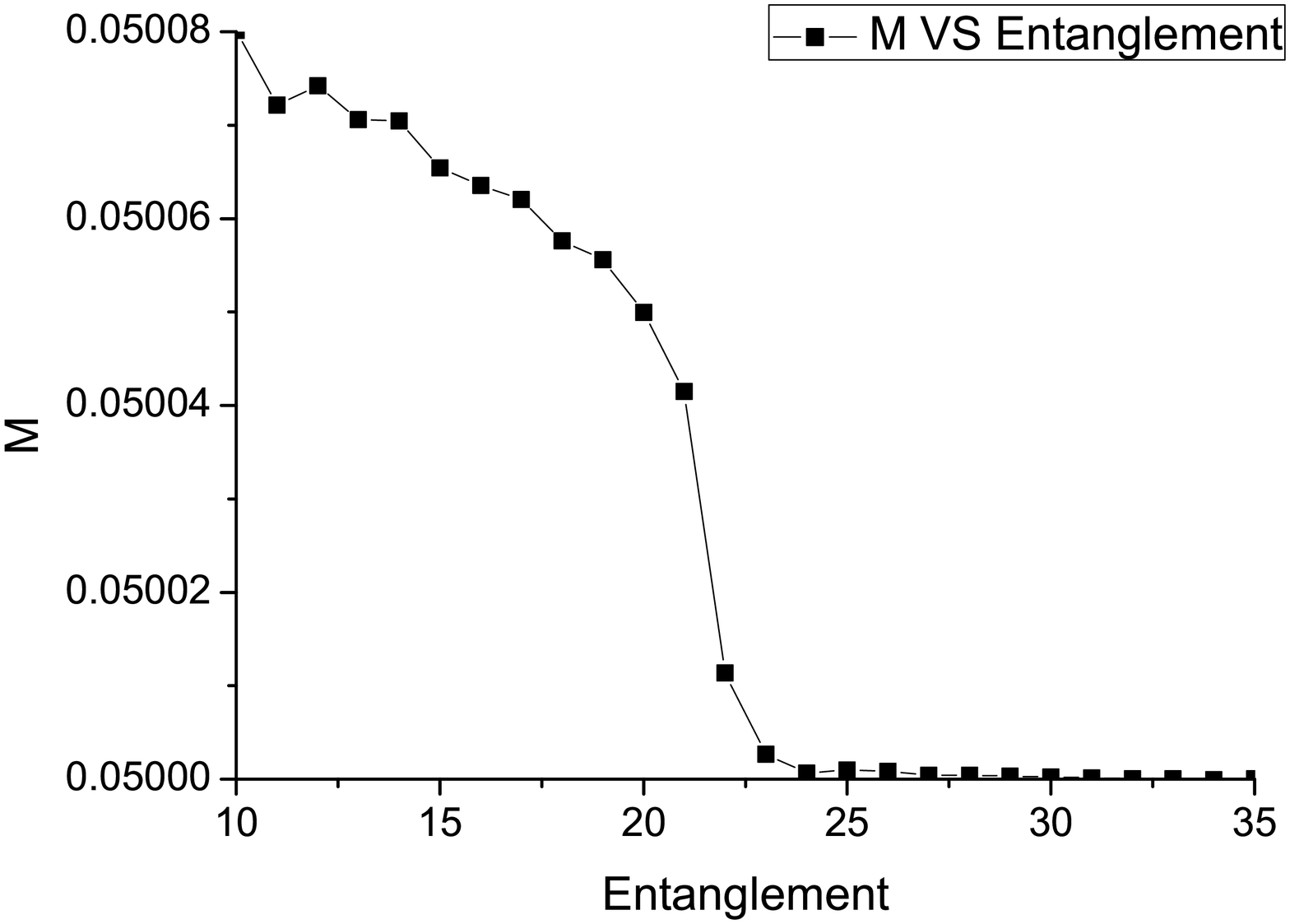}
\caption{\label{fig:MVSP}M converges to 0.05 suddenly at entanglement 24. $B=0.1348$ and $\lambda=0.5$} 
\end{center}
\end{figure}

For the moment, let us consider the Bosonization's prediction on the xxz model in external magnetic field $B$, where the bosonization makes precise prediction for both 2-point and 4-point functions since the asymptotic logarithmic correction is absent for non-zero $B$. The 2-spin correlation function is predicted to be \cite{Hikihara},
\begin{align}
\label{twozz}
\langle S_l^zS_{l+r}^z\rangle = & M^2-\frac{1}{4\pi^2\eta r^2}\notag\\
& +\sum_{n=0}^{\infty}{A_{2n+1}^z\left( -1\right)^r\frac{cos\left[\left(2n+1\right)Qr\right]}{ r^{\left(2n+1\right)^2/\eta}}}
\end{align}
where the decay exponent $\eta$ is a function of $\lambda$ and the magnetization $M$,
$A_n^z=\frac{a_n^2}{2}$ and $Q=2\pi M$ is an incommensurate wave number.
The 4-spin correlation functions are likewise given by \cite{Hikihara},
\begin{align}
\label{fourcorrelation1}
& \langle :\left(S_l^+S_{l+1}^-+S_l^-S_{l+1}^+\right)::\left(S_{l+r}^+S_{l+r+1}^-+S_{l+r}^-S_{l+r+1}^+\right):\rangle\notag\\
= & 16B_1\frac{\left(-1\right)^r}{ r ^{\frac{1}{\eta}}}\mathrm{cos}\left(Qr\right)+\frac{16B'_0}{ r^2}+\cdots
\end{align}
\begin{align}
\label{fourcorrelation2}
\langle S_l^+S_{l+1}^+S_{l+r}^-S_{l+r+1}^-\rangle=\frac{8B_2}{ r^{4\eta}}+\cdots
\end{align}
where $:\Re: \equiv \Re- \langle \Re \rangle$. In principle, the above coefficients
$B_1$, $B'_0$, $B_2$, $a_n$ and so on can be determined by fitting to numerically computed
correlation functions. 
Authors of \cite{Hikihara} indeed used DMRG to determine these quantities.
We have calculated the two and four spin correlation functions for
the anisotropy $\lambda=0.5,0,-0.5$, and the magnetization $M=0.05,0.25$ for a 200-sites 
spin-$\frac{1}{2}$ chain, 
as \cite{Hikihara} did. The difference is that we used the periodic boundary condition while 
\cite{Hikihara} used the open boundary condition which is more accurate than
the periodic boundary condition in the DMRG calculation. 
We vary the magnetic field $B$ to gradually approach an aimed $M$ for each given $\lambda$. 
Tab.\ref{table:MBlambda} gives the magnetic field intensities for some combinations of $M$ and $\lambda$. 
\begin{table}
\begin{center}
\begin{tabular}{|c|c|c|c|}
  \hline
  B  & $\lambda=-0.5$  & $\lambda=0$ & $\lambda=0.5$  \\
  \hline
  M=0.05 & 0.03395 & 0.07820 & 0.13480 \\
  \hline
  M=0.25 & 0.16058 & 0.35350 & 0.56680 \\
  \hline  
 
\end{tabular}
\caption{\label{table:MBlambda}The magnetic field intensities for different combinations of $M$ and $\lambda$.}
\end{center}
\end{table} 
In fact, one can vary $B$ to have the calculated $M$ close to the aimed value. This can be done quickly 
at small entanglement. Then a convergence is checked with respect to increasing entanglement. One sees that the 
calculated $M$ suddenly converges to the aimed value in Fig.\ref{fig:MVSP}.  
\begin{figure}
\begin{center}
\includegraphics[height=2.8in, width=3.2in]{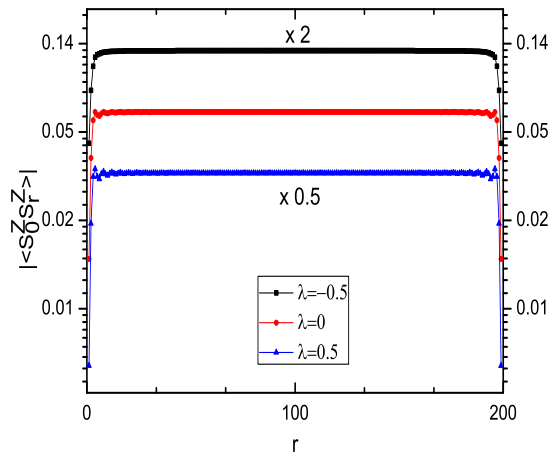}
\caption{\label{fig:zzcorrelation}$\langle S_0^zS_r^z\rangle$ by EPT does not have the artificial oscillation at the edges.} 
\end{center}
\end{figure}   
For the 2-spin correlation, we mainly compare $\langle S_0^zS_r^z\rangle$ with (b) of Fig.1. in 
\cite{Hikihara}, while $\langle S_0^xS_r^x\rangle$ is quite similar to (a) of that figure 
(we use the periodic boundary condition). Fig.\ref{fig:zzcorrelation} shows that 
$\langle S_0^zS_r^z\rangle$ by EPT, unlike DMRG, does not have the artificial oscillation at the edges. 
\begin{figure}
\begin{center}
\begin{tabular}{c}
(a) $\lambda=0.5,M=0.05,L=200$\\
\includegraphics[height=2.5in, width=3.in]{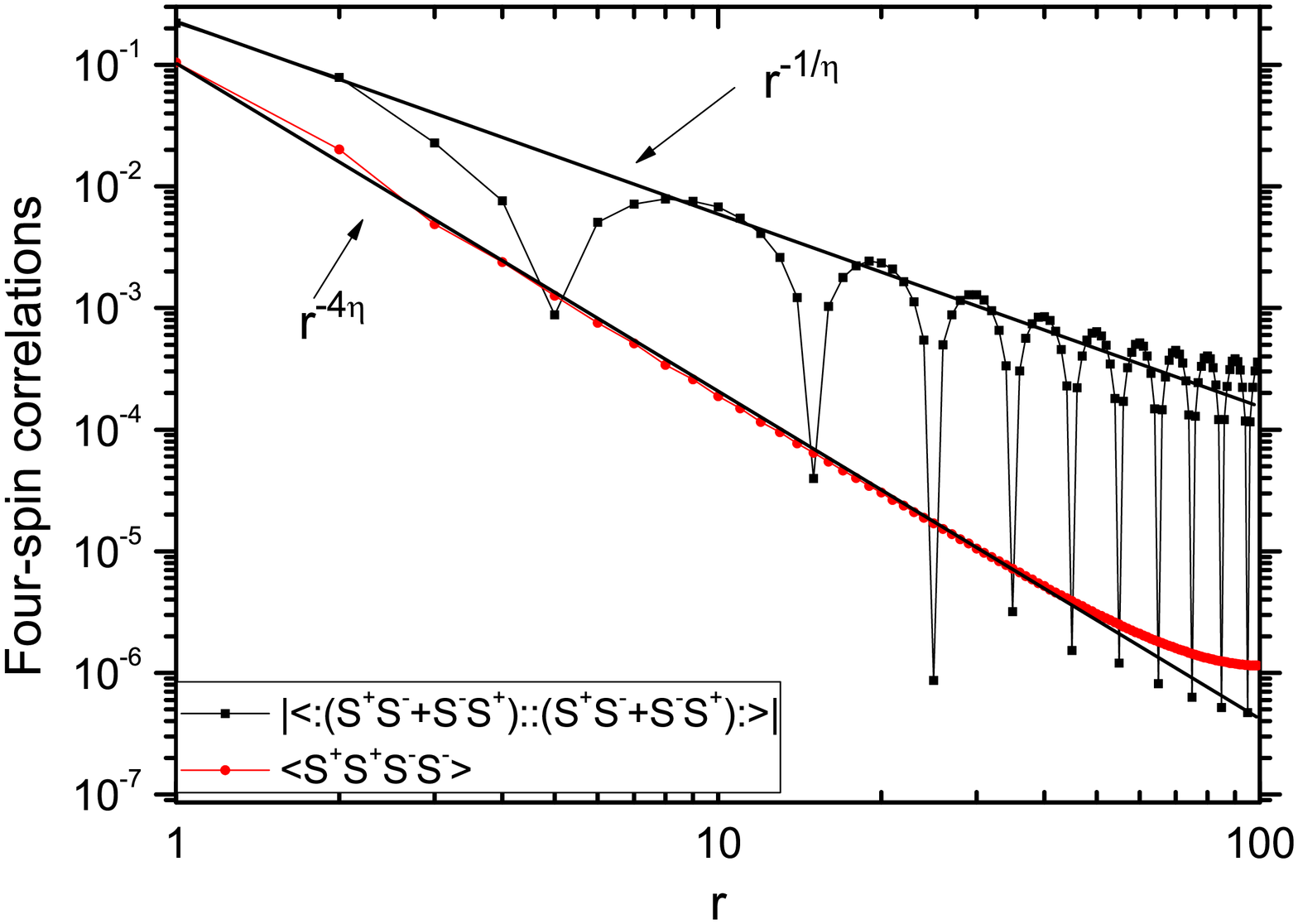}  \\
(b) $\lambda=0,M=0.05,L=200$\\
\includegraphics[height=2.5in, width=3.in]{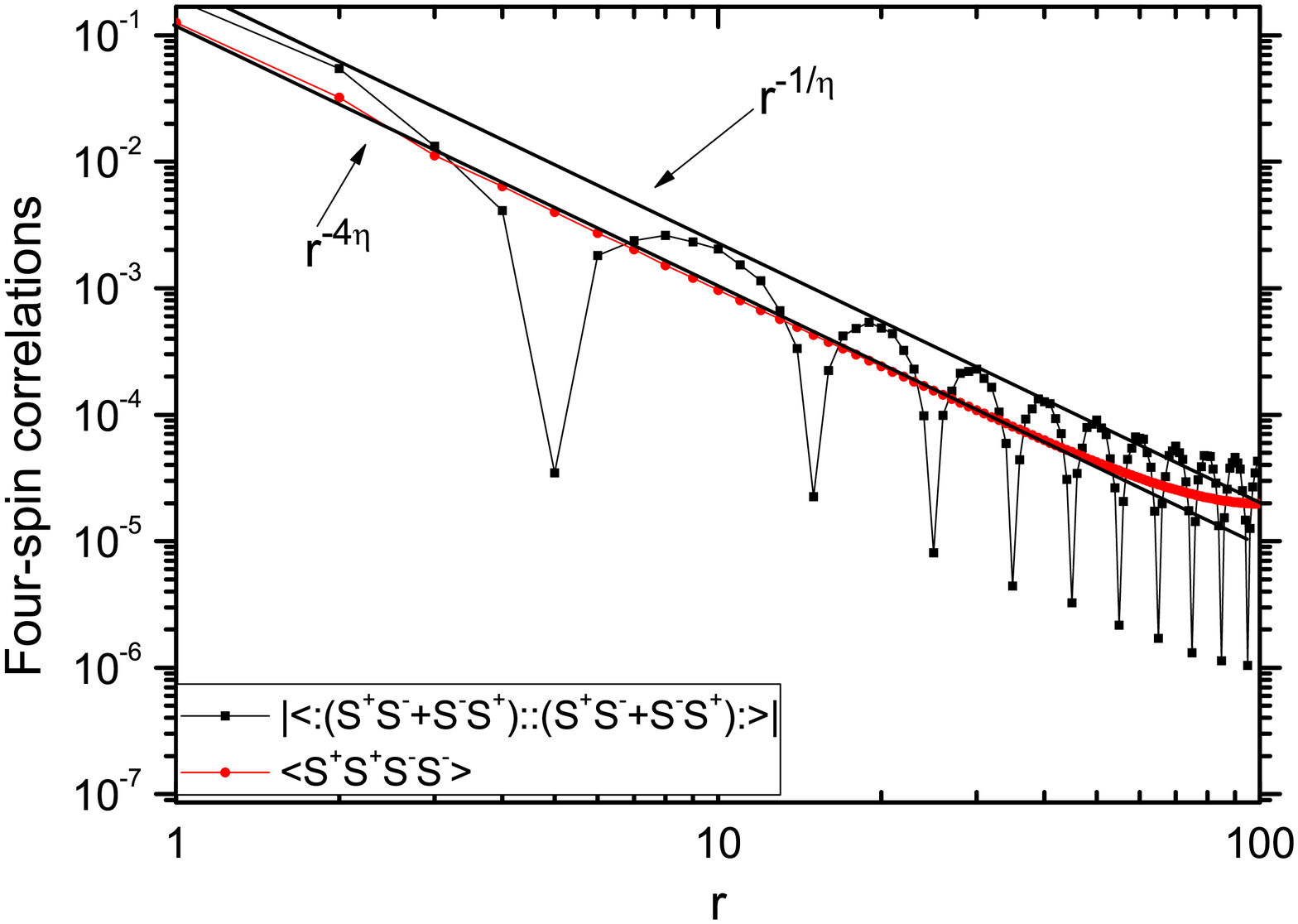} \\
(c) $\lambda=-0.5,M=0.05,L=200$\\
\includegraphics[height=2.5in, width=3.in]{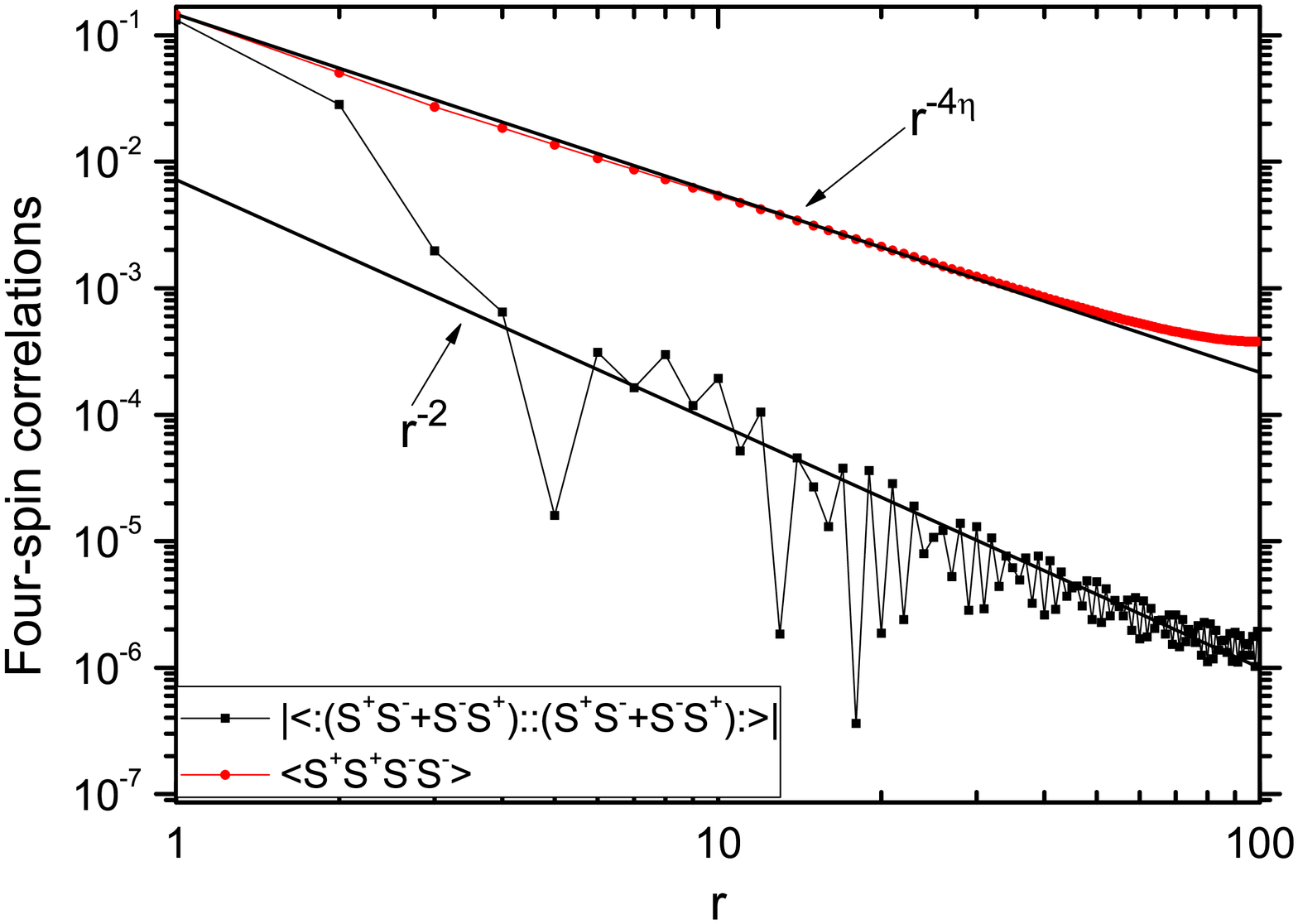}
 \end{tabular}
\caption{\label{Fig:fourspin}The four-spin correlation functions for $M=0.05,\lambda=-0.5,0,0.5$.}
\end{center}
\end{figure} 

We now present in Fig.\ref{Fig:fourspin} the 4-spin correlation functions by EPT for $\lambda=-0.5,0,0.5$ 
and $M=0.05$. The calculations are all converged with the entanglement. A small deviation
near the center of the chain from the power-law decay is due to the periodic boundary condition.   
$\langle S_l^+S_{l+1}^+S_{l+r}^-S_{l+r+1}^-\rangle$ and the envelope 
of $\langle :\left(S_l^+S_{l+1}^-+S_l^-S_{l+1}^+\right)::
\left(S_{l+r}^+S_{l+r+1}^-+S_{l+r}^-S_{l+r+1}^+\right):\rangle$ decay in a power-law as a
 linear fitting in a log-log plot shows, which is predicted by \cite{Hikihara}. However, we see a clear 
period-10 oscillation in plots  $(a)$ and $(b)$ while DMRG does not show this (Fig.3. in \cite{Hikihara}). 
In fact, plot $(c)$ also has an oscillation with the period of 10, although subtracting the mean value from 
the correlation makes it hardly visible. This oscillation was clearly seen for the first time owing 
to EPT's great precision. Note that $Q=0.1\pi$ when $M=0.05$ and the absolute values were taken in the plots. It is exactly what was predicted by the term $\mathrm{cos}(Qr)$ in (\ref{fourcorrelation1}). 

\subsection{Comparison with CFT}

Because of a required high precision, CFT's prediction of logrithmic correction to a power law decay of spin-spin correlations has not been addressed by any methods so far, to our best knowledge. 
In Fig.12, we plot $W_z$ and confirmed the CFT result. Note that the convergence of $W_z$ with 
entanglement to CFT is seen around $r \sim 1000$ for the entanglement $p \geq 180$.  
This also shows that an asymptotic behavior is realized for $r \sim 1000$. 
At this extreme accuracy, the obtained ground state energy -0.4431467 is very close to 
the exact value of Bethe ansatz -0.4431471. It would be interesting to check another 
prediction of CFT \cite{affleck} and series expansion \cite{fisher}: 
if and how the asymptotic correlation functions exhibits critical behavior 
at $\lambda=1$


\begin{figure}
\begin{center}
\includegraphics[height=2.8in, width=3.2in]{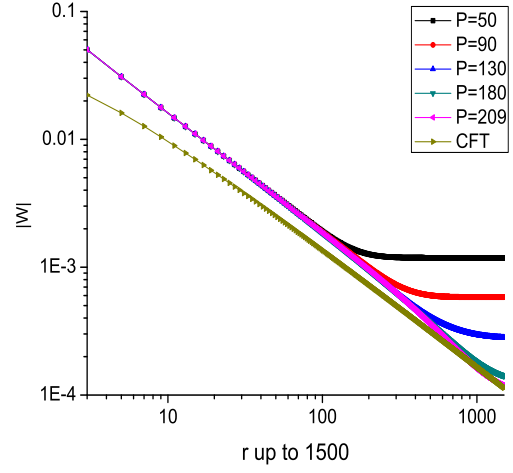}
\caption{\label{fig:infinite}Spin-spin correlation functions for an infinite spin-$\frac{1}{2}$ chain}
\end{center}
\end{figure}

\section{\label{one}Spin-1 chains}

\begin{figure}
\begin{center}
\includegraphics[height=2.8in, width=3.2in]{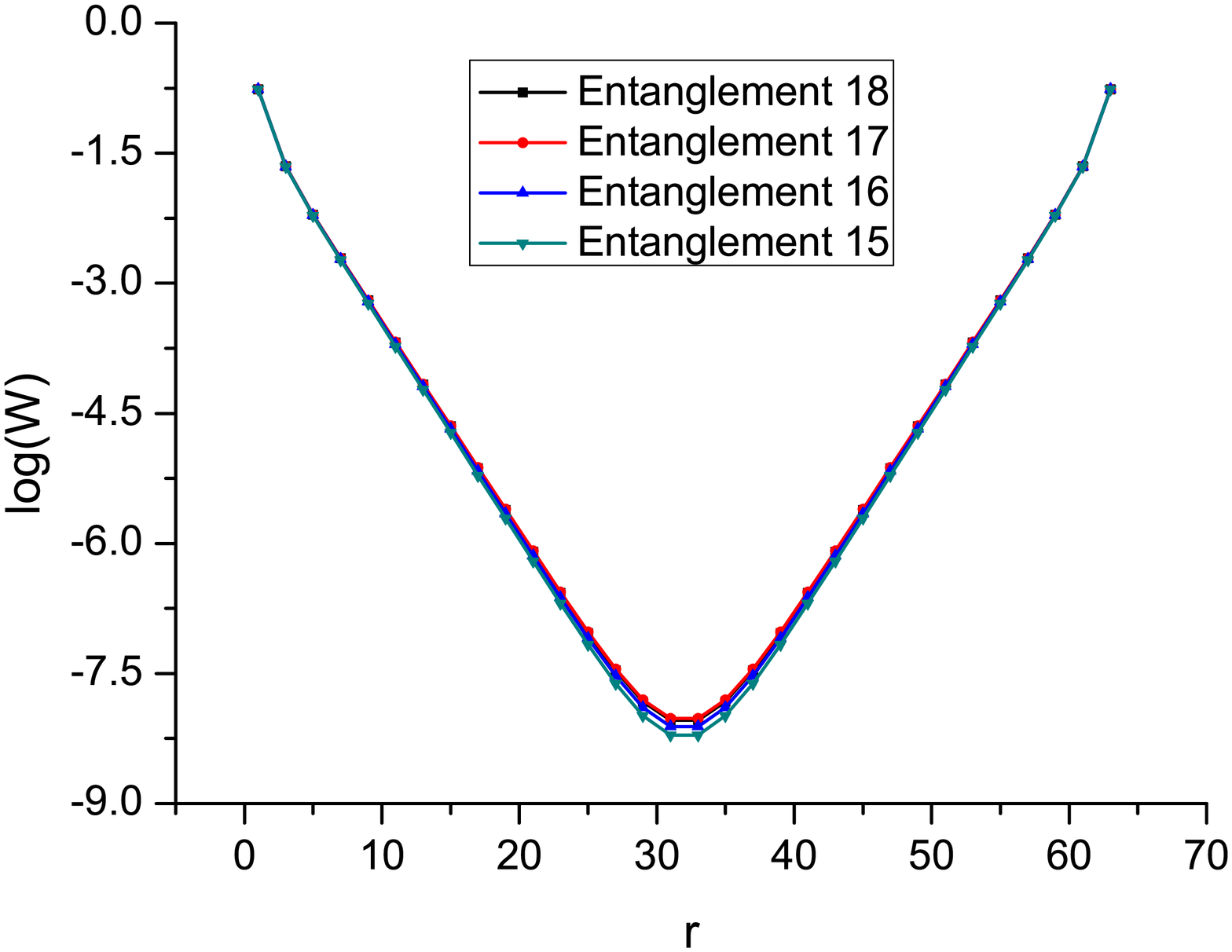}
\caption{\label{fig:onecorr}The spin-spin correlation functions exponentially decay with the distance} 
\end{center}
\end{figure}

Similarly we calculated the spin-1 chains of many different lengths. We compare the EPT results with DMRG. 
The GS energy for a 48-sites chain is -1.401482 by EPT at entanglement $p=20$ versus -1.401484 by DMRG \cite{white1}. 
The energy gap between the GS and the 
first excited state in the same calculation is 0.41242 by EPT and 0.41232 by DMRG. 
The spin-spin correlations were calculated from the GS wave function. We see a clear convergence in the semi-log 
curve of the correlation functions even as early as at $p=18$ for a 64-sites chain in Fig.\ref{fig:onecorr}. 
The straight line in the semi-log plot indicates an exponential decay of the spin-spin correlation with the 
distance, consistent with the existence of the energy gap in spin-1 chains.  
  
We have also calculated the GS phase diagram of the spin-1 xxz chain,  
 $\lambda\neq 1$, using the Roomany-Wyld RG finite-size scaling \cite{roomany}.
\begin{figure}
\begin{center}
\includegraphics[height=2.8in, width=3.2in]{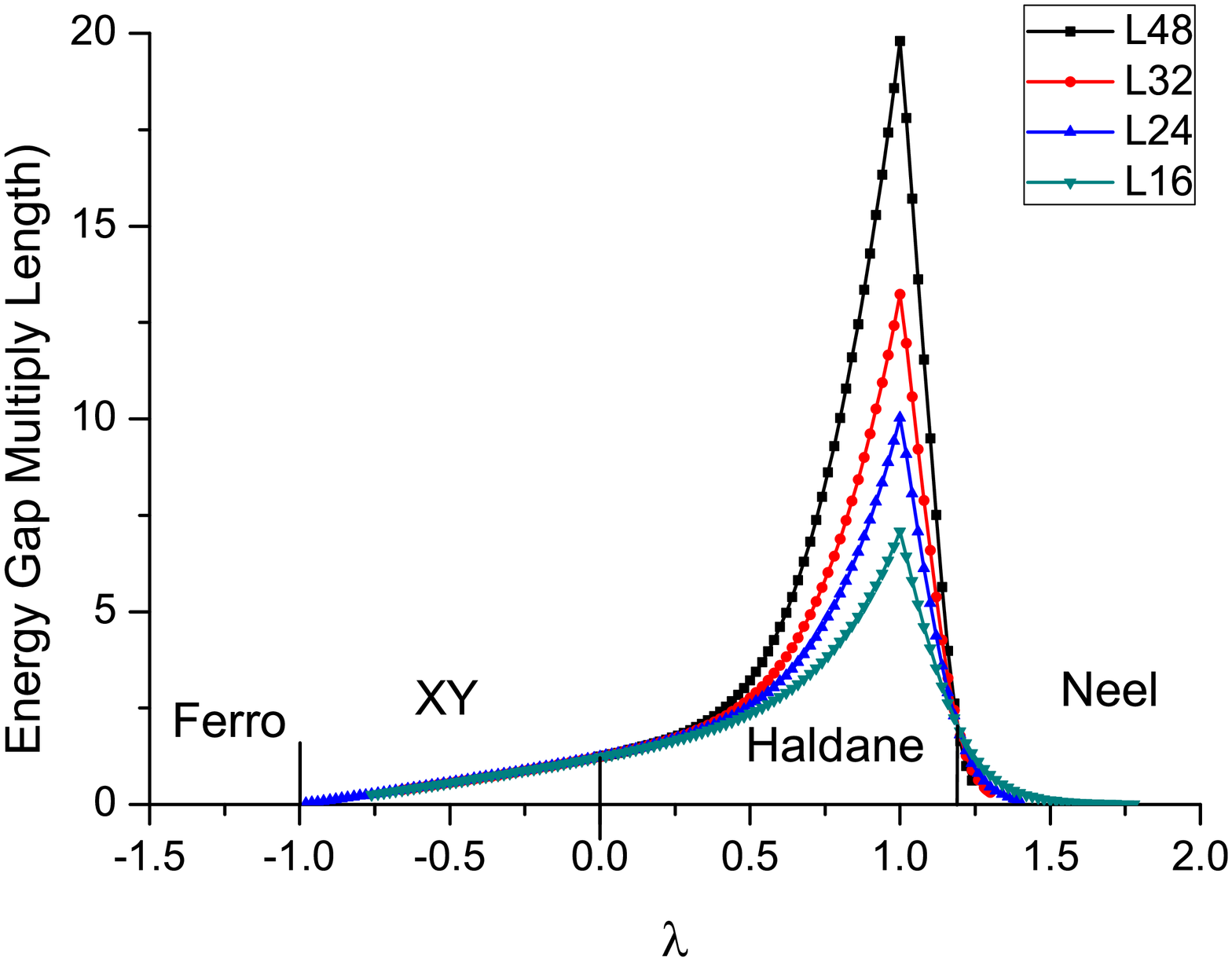}
\caption{\label{fig:phase1}Phase diagram of the spin-1 xxz chain} 
\end{center}
\end{figure}  
In Fig.\ref{fig:phase1}, the four phases are identified in the plot of Gap(L) x L,
where Gap(L) denotes the energy gap as a function of the chain size L. When $\lambda$ is fairly 
large, it approaches the Ising limit. The GS is a unique Neel state. 
The spin-1 xxz chain enters from Neel phase to Haldane phase at $\lambda=1.192$, 
where the energy gap disappears. 
From this point to $\lambda=0$ there is an energy gap for all chain length. 
And at $\lambda=1$, the energy gap converged to a finite value, the Haldane gap. 
From $\lambda=0$ to $\lambda=-1$ is the XY phase, the infinite chain has no energy gap. At $\lambda=-1$, all finite chains have no energy gap.
After this point, it enters ferromagnetic phase, where the GS is doubly degenerate. 
Our phase diagram agrees with that of
the level-spectroscopy method \cite{wei}.  

\section{\label{sec:conclution}Conclusion}
Two EPT algorithms, EPT-g1 and EPT-g2, have been applied to calculate the ground sate properties 
of AF spin chains. They give exactly the same results with the same 
convergence speed. Among them, EPT-g1 is especially suitable for the infinite system. 
By successfully comparing with Bethe ansatz, CFT and Bosonization, we have seen the EPT's ability to
handle quantum systems of arbitrary size.  This is especially 
a good news to field theory because some of their predictions have not been verified by any numerical calculations before. 
Along with its further ability to calculate excited states either by EPT-g algorithm or EPT-e \cite{chung5},
EPT's broad applicability in 1D quantum systems will be clear.  

Among many possible applications and desirable extensions of EPT, we are currently examining the non-uniform matrix product states 
in the spin-$\frac{1}{2}$ AF Heisenberg models in 1D and 2D. 
In particular, the $J_1-J_2$ model in 2D including the square and triangular lattices. Since such a non-uniform system is
more or less inherently of short-range nature, one would not expect much improvement over DMRG.  A question is rather 
if the non-uniform EPT can still maintain its simplicity and stability in algorithm and numerical efficiency. A work
is currently underway, and will be reported in the near future.  
A true new aspect, however, might emerge when we consider impurities/nano-structures embedded in strongly correlated host 
materials of macroscopic extension, possibly realized in Kondo phenomena.  It is quite interesting to see how EPT works here.    


\begin{acknowledgments}
S.G.C thanks Prof. Kenn Kubo for helpful discussions. This work was supported by a grant from the Faculty Research and Creative Activities Support Fund, WMU. It partially utilized the College of Sciences and Humanities Cluster at the Ball State University and the RICC at Advanced Center for Computing and Communication, RIKEN. 
\end{acknowledgments}


\bibliography{spin_chain}

\end{document}